# Near-Field Plasmonic Behavior of Au/Pd Nanocrystals with Pd-Rich Tips

*Emilie Ringe,*[1*] *Christopher J. DeSantis,*[2] *Sean M. Collins,*[3] *Martial Duchamp,*[4] *Rafal E. Dunin-Borkowski,*[4] *Sara E. Skrabalak,*[2] *Paul A. Midgley*[3]

1. Department of Materials Science and NanoEngineering, Rice University, 6100 Main St., Houston TX 77005, USA

2. Department of Chemistry, Indiana University, 800 E. Kirkwood Ave., Bloomington, IN 47405, USA

3. Department of Materials Science and Metallurgy, University of Cambridge, 27 Charles Babbage Road, Cambridge CB3 0FS, UK

4. Ernst Ruska-Centre for Microscopy and Spectroscopy with Electrons (ER-C) and Peter Grünberg Institut 5 (PGI-5), Forschungszentrum Jülich GmbH, D-52425 Jülich, Germany

TOC FIGURE

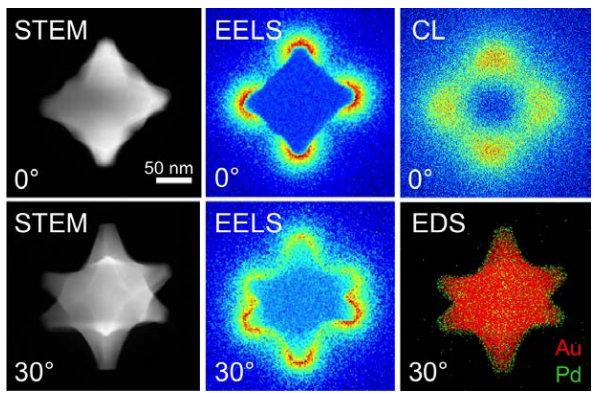

ABSTRACT

Using nanometer spatial resolution electron-energy loss spectroscopy (EELS), energy dispersive X-ray spectroscopy (EDS), and cathodoluminescence (CL) mapping, we demonstrate that Au alloys containing a poor plasmonic metal (Pd) can nevertheless sustain multiple size-dependent localized surface plasmon resonances and observe strong field enhancement at Pd-rich tips, where the composition is in fact least favorable for plasmons. These Au/Pd stellated nanocrystals are also involved in substrate and interparticle coupling, as unraveled by EELS tilt series.

MANUSCRIPT TEXT

Nanoparticles of plasmonic metals such as Au, Ag, Cu, and Al can sustain narrow and intense localized surface plasmon resonances (LSPRs), light-driven coherent oscillations of the conduction electrons. LSPRs and related phenomena are utilized in a variety of fields, from biological sensing to cancer therapy.[1,2] By incorporating dielectric shells, layering different metals or creating heterogeneous aggregates, applications have been extended to, for example, surface-enhanced Raman scattering (SERS) tags and hydrogen sensors.[3-5] Multi-metal nanoparticles are of tremendous interest, as they provide routes to multifunctional behavior and novel properties; coating or incorporating a catalytic component on a plasmonic nanoparticle enables in-situ reaction tracking, the creation of novel attachment chemistry as well as the possibility of plasmon-enhanced catalysis and hot electron injection.[6-9] However, several catalytic metals such as Pd and Pt are poorly plasmonic, displaying broad and heavily damped resonances.[10-12] The incorporation of these poor plasmonic metals in a multifunctional system, while catalytically desirable, could be detrimental to plasmon resonaces, unlike in the well-studied AgAu plasmonic alloys and core-shell systems.[13-15] Far-field optical studies have shown LSPR signatures as well as field enhancement via SERS in Pd-containing alloy nanoparticles.[16-22] Here, we further explore the effects of Pd in plasmonics by presenting a detailed, spatially resolved study of the composition and electric field distribution of alloys containing a poor plasmonic metal, Pd. We demonstrate that particles incorporating a catalytically active but heavily damped material can sustain multiple size-dependent LSPRs that are narrow and strongly localized at the Pd-rich tips, and can couple with a dielectric substrate as well as other nanoparticles. By establishing that the full range of plasmonic characteristics expected of Au is maintained in Au-Pd alloys, this study, the first of its kind to the best of our knowledge, establishes the feasibility and provides strong motivation for further research in multifunctional plasmonic-catalytic systems.

Stellated Au/Pd nanocrystals were synthesized according to a previously reported colloidal method in which Au and Pd precursors are co-reduced to deposit metal onto octahedral Au seeds.[21] Overgrowth in the <111> directions creates 8-branched structures with point group symmetry $O_h$ called octopods. The protrusions forming the 8 branches are well-defined and terminated by flat {111} facets, as is evidenced by the sharp edges present in the scanning transmission electron microscopy (STEM) images (Figure 1) as well as the electron tomograms obtained from high angle annular dark field (HAADF)-STEM tilt series for over 10 particles (examples in Supporting Information). This stellated external morphology is, rather surprisingly, formed by a twin-free, single crystalline nanoparticle, as evidenced by nanometer resolution diffraction mapping, of which a few snapshots are shown in Figure 1. The convergent beam electron diffraction pattern displays the same symmetry and orientation for any position of the sub-nanometer probe, revealing that each region of the crystal has a similar crystallographic orientation. This finding, reproducible for all the particles observed, indicates that the single crystalline nature of the seeds is conserved through the synthesis. The fully miscible Au and Pd form a continuous solid solution through the tips of the particle, rather than a patchy or polycrystalline core-shell structure, as shown by atomic resolution imaging (Figure 1 and Supporting Information). Diffraction patterns, Fourier transforms of lattice images, and atomic spacing measurements all yield a lattice spacing between that of Au and Pd, i.e. between 408 and 398 pm, also consistent with a solid solution, within the measurement error of a few percent. This lattice continuity, surface smoothness, and lack of scattering defects in Au/Pd particles are important attributes as they likely have a favorable impact on the quality and lifetime of the plasmon resonances.[23]

The Pd in octopods is mainly concentrated at the tips of the branches, where the electric field intensity is highest (*vide infra*), as evidenced by STEM-energy dispersive X-Ray spectroscopy (EDS). Easily interpretable STEM-EDS maps were obtained by tilting the sample 30° such that most (6/8) of the branches were isolated and directly addressable (Figure 1). In this configuration, the Pd segregation appears in both images and linescans as an increase in the relative Pd X-ray intensity at the very tip of the branches. Relative X-ray intensity is defined here as the background subtracted integrated peak intensity divided by the total X-ray signal; this normalization is essential to decouple the effects of thickness and compositional changes. Such results are reproducible and expand on recently published findings,[19,20,22,24] confirming the unique composition profile of Au/Pd octopods. The synthetic method, seed-mediated co-reduction, is likely the origin of the very steep gradient of Pd at the tips as Au deposits at a faster rate than Pd on the Au-seeds and is thus depleted as the growth proceeds, a process also observed with Pt alloys.[25,26] Post-synthesis segregation is not expected thermodynamically since the surface free energy of Pd is higher than that of Au, at least in a clean environment.[27]

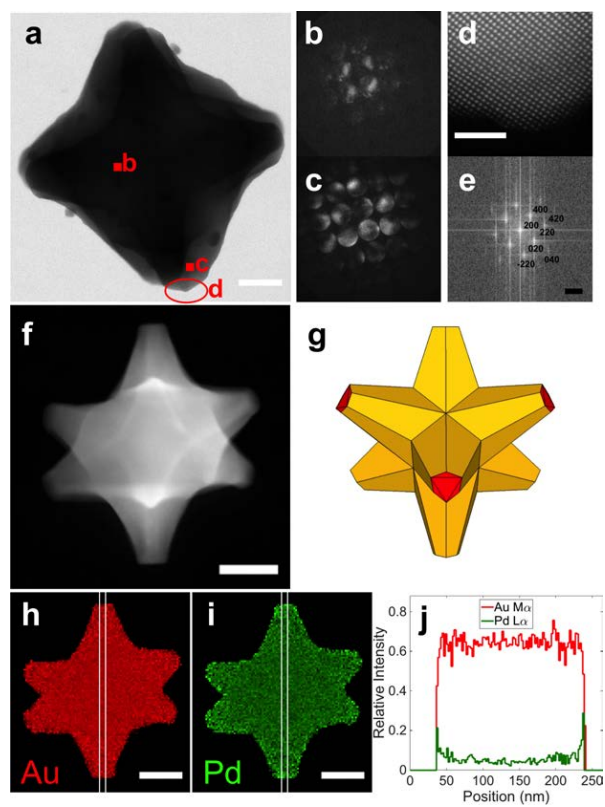

**Figure 1.** Structure, crystallography, and composition of Au/Pd octopods. (a) STEM bright field image. (b-c) Convergent beam electron diffraction patterns obtained at the positions indicated in a, <001> orientation. (d-e) HAADF-STEM image of the region shown in a and associated Fourier transform, <001> orientation. (f) HAADF-STEM image of an octopod tilted -30°. (g) Model of an octopod tilted -30°. (h) Au $M_\alpha$, $L_\alpha$, and $L_\beta$ summed relative X-ray intensity map. (i) Pd $L_\alpha$, $L_\beta$, and $K_\alpha$ summed relative X-ray intensity map. (j) Relative X-ray intensity linescan of the Au $M_\alpha$ and Pd $L_\alpha$ lines along the vertical axis in h-i. Scale bars, 25 nm for a, 2 nm for d, 5 $nm^{-1}$ for e, 50 nm for f, h, i.

Far-field studies, probing scattering, absorption or extinction, are intrinsically diffraction-limited, lacking the spatial resolution needed to understand field distribution and mode symmetry; this important information can be obtained by near-field techniques that provide the intensity and distribution of electromagnetic field enhancement around a particle. Results from near-field studies can direct synthesis, functionalization and substrate immobilization strategies as well as define optimal architectures for applications ranging from sensing to plasmon-enhanced catalysis.[8,28] STEM recently emerged as a powerful tool to probe simultaneously size, composition, shape, and local plasmon resonance distribution in nanoparticles.[29-40] Indeed, a high energy electron beam can interact with plasmons or LSPRs; the magnitude and energy of this interaction can be probed either by tracking the energy lost by electrons *via* electron energy loss spectroscopy (EELS) or by collecting the light emitted as a result of plasmon decay *via* cathodoluminescence (CL) spectroscopy.[37-39,41,42] Simple shapes such as triangles, rods, shells, icosahedra and cubes[30,32,36,38-40,43-48] have been studied, albeit to the best of our knowledge the near-field of alloy nanocrystals containing a poor plasmonic metal has never mapped with nanometer resolution as is done here with STEM-EELS and CL.

Stellated nanoparticles such as octopods are of particular interest to sensing and surface-enhanced spectroscopies as their sharp tips concentrate the electric field and lead to intense and high refractive index sensitivity LSPRs.[21,49,50] The presence of Pd at the tips, evidenced by EDS, provides catalytic sites interesting for plasmon-enhanced catalysis and in-situ sensing, however Pd is a heavily damped plasmonic material not suitable for efficient optical excitation when in pure form, and, presumably, in Pd-rich alloys. Resonances in Au/Pd octopods have been studied optically, showing high extinction coefficient and high refractive index sensitivity,[21] motivating further studies and clear demonstration of the effect of Pd alloying in the near-field. The results presented here show that Pd-rich tips do not prevent strong localized near-field intensity.

The plasmon behavior of 11 isolated octopods and several aggregates was analyzed using STEM-EELS and STEM-EELS tilt series, representative spectra from a particle at 0° and +30° acquired on a monochromated FEI Titan Themis are presented in Figure 2. When the sub-nanometer beam is far from the particle, the signal obtained is simply the tail of the energy spread of the incoming electron (i.e. the tail of the zero-loss peak, ZLP), while spectra obtained with the beam close to particle tips display a broad but intense feature in the 1.5-3 eV range. Reconstructed energy filtered (EFTEM) images obtained around the peak of the main feature in single and aggregated nanoparticles provide an indication that the field intensity is concentrated at the tip of the branches and that nearby particles can interact (Figure 2). Contributions from the tail of the ZLP and the spectral and spatial overlap of high order modes make direct analysis of the raw spectra or reconstructed EFTEM images of limited use, however. To overcome this difficulty, we used a powerful blind source separation technique (non-negative matrix factorization) to extract individual plasmon resonances in single octopods and de-couple the effects of the excitation energy spread, examples of results are reported in Figure 3 and in Supporting Information.[30,51]

Mapping plasmon modes in an electron microscope is inherently a correlated structure/function measurement, as electron-beam images from the HAADF and/or dark field (DF) detectors can be acquired concurrently with the EELS spectrum image (SI). Thus, the relationship between mode energy and particle size is readily addressable, we report it in Figure 2 for the lowest energy mode, a dipolar resonance associated with fields penetrating inside the substrate (*vide infra*). The peak position considered here is the maximum of the decomposed spectral component corresponding to this resonance, rather than the maximum of the entire EELS spectrum. This distinction is very important: the maximum of the later is inevitably shifted due to the ZLP and overlapping resonances, while the maximum of a decomposed resonance is real as it only contains contributions from this specific LSPR. As expected from retardation effects and as observed for other shapes,[52-55] the resonance energy decreases with increasing particle size. In a previous report using aqueous bulk solution extinction, the maximum of the spectrum of Au/Pd octopods was observed to vary nearly linearly from 565 to 746 nm (2.194 to 1.662 eV) when the average distance between opposing branches increased from 61 to 143 nm. EELS shows a decrease from 2.1 to 1.5 eV when the particle size increases from 106 nm to 282 nm, an agreement in trend with an offset justifiable on the basis of the different refractive index environments. Indeed, the blue shift observed in vacuum/$Si_3N_4$ membrane (EELS) compared to water (bulk extinction measurements[21]) is equivalent to that caused by a change in refractive index of 0.16 refractive index units, as calculated from the refractive index sensitivities and LSPR energies obtained optically.[21] A refractive index of ~1.17 (1.333 for water minus 0.16) is well justifiable: vacuum and $Si_3N_4$ have refractive indices of 1, and ~2.05, respectively, and the particle interacts mostly with the former, given only the tips of four branches are touching the substrate. Other effects could play a role but appear to only minimally affect the energy shift in this system, namely the size- and shape-dependence of the refractive index sensitivity, the error associated with determining the position of the plasmon resonance in bulk extinction measurements due to higher order modes, and the expected small redshift of the near-field peak with respect to far-field maximum.[56,57] While the correlation in Figure 2 is rather linear, two particles (229 and 240 nm size) with typical octopod shapes appear as outliers, most likely due to the heavy oxygen/argon plasma cleaning they uniquely were subject to, shifting the refractive index downwards (hence the plasmon energy upwards)[58] by removing any remaining carbonaceous surfactants from the surface of the particle. Overall, this correspondence between EEL and extinction spectroscopy shows that the expected energy shifts are observed even when only the Pd-rich tips are in contact with a high refractive index environment, indicating that the very ends of the branches are capable of local sensing, a conclusion not directly achievable optically.

The plasmonic behavior of dimers and aggregates proves that the Pd-rich tips also support interparticle coupling. Simulated EFTEM images from STEM-EELS data acquired on a JEOL-ARM microscope (Figure 2) clearly show the dominant bright bonding mode at 1.5 eV and the dark antibonding mode at 2.3 eV, respectively lower and higher in energy than the expected LSPR energy for single octopods of this size (~150 nm, LSPR~1.8-2.0 eV).

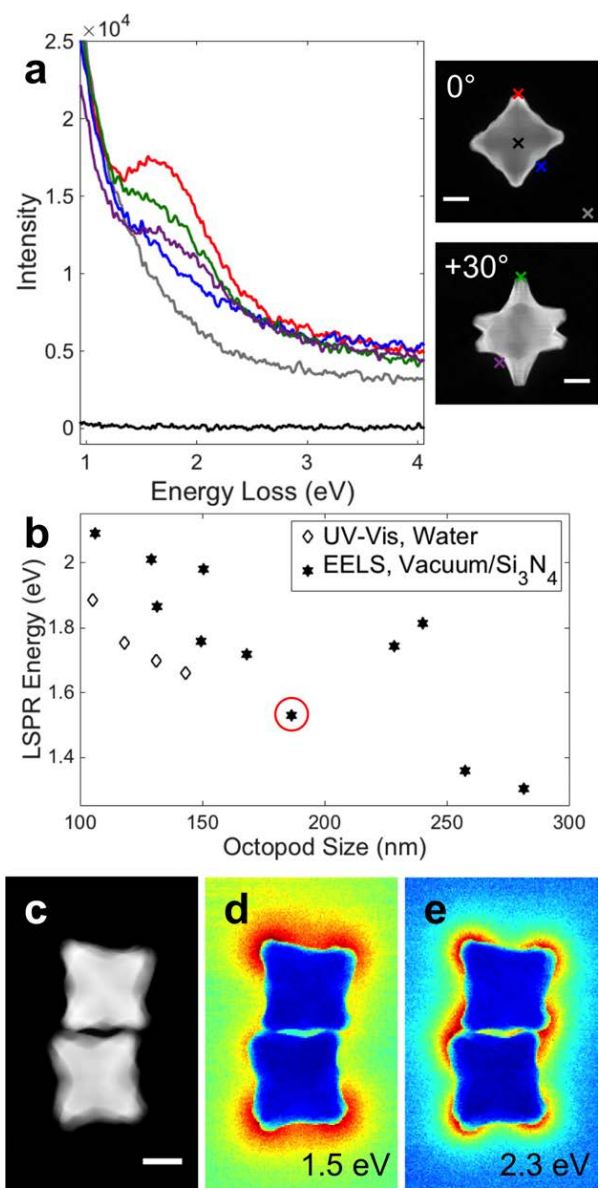

**Figure 2.** EELS of Au/Pd octopods, size dependence of the LSPR energy and plasmon coupling. (a) Monochromated STEM-EEL spectra at the different positions marked on the right HAADF-STEM images for the same particle tilted 0° and +30°. (b) Correlation between size (face diagonal) and energy of the lowest energy LSPR; the particle shown in a is circled in red. UV-Vis data from ref.[21] (c) HAADF-STEM image of an octopod dimer. (d-e) Reconstructed EFTEM images with a 0.2 eV slit centered at 1.5 and 2.3 eV. Scale bars, 50 nm.

The near-field plasmonic behavior of Au/Pd nanocrystals was studied by performing a statistical analysis (NMF) on STEM-EELS tilt series, an example of which is presented in Figure 3 (additional tilt series in Supporting Information). This particle was confirmed to contain a increased Pd concentration at the outer edges by EDS (Figure 1 and Supporting Information). The SI as a whole was fit with NMF, extracting spectral factors that include the tail of the ZLP, plasmon modes, and other energy loss contributions. Each of those factors are multiplied at each pixel by unique, position-dependent loadings to generate a global fit; a map of the loadings provide the spatial distribution, i.e. contribution to the overall EELS intensity, of a given spectral factor. The fit matches the raw data very well, showing that the 6 spectral factors extracted fully explain the observed spectral variations in the SI. In Figure 3a, two narrow spectral factors (2 and 3) can be attributed to plasmon resonances. These LSPRs appear to be hybridized with the substrate analogously to the proximal and distal plasmon modes of silver nanocubes.[30,54,59] The proximal LSPR (2) is consistent with a field distribution penetrating into the higher refractive index substrate, and is thus lowest in energy, between 1.3 and 2.1 eV as shown in Figure 4. The distal LSPR (3) has high field intensity predominantly away from the substrate, penetrating into the low refractive index of the surrounding vacuum; this mode is consequently higher in energy, above 2 eV.

This coupling with the substrate is obvious from the tilt series (Figure 3 and Supporting Information): as the particle is tilted, some of its branches are moved away from being superimposed with the branches touching the substrate and strongly and uniquely contribute (rather than being shadowed or averaged) to the high energy distal resonance observed (LSPR 3). These branches are towards the top of Figure 3 for positive tilts and the bottom for negative tilts. In this experiment a positive increase in tilt corresponds to the top of the substrate moving away from the reader (and the electron gun). As expected, the proximal LSPR (2) behaves inversely, that is, intensity is higher in the region (bottom at positive tilts) where the substrate is closer to the reader. This finding confirms that even when a poor plasmonic metal such as Pd is present at the tips of particles, strong localized fields are sustained and interact with the substrate. Note that the two LSPR features, while observed in similar orientations for all particles studied, vary in relative intensity. We have not yet been able to fully elucidate and quantify this relationship; it appears that the particle size, the area of contact with the substrate, the shape and size of the branches, and the asymmetry within a given particle all might play a role. Additional spectral factors, beyond the proximal and distal modes, are extracted by NMF, but are not as relevant for plasmonics. Briefly, spectral factor 1 is the tail of the zero-loss peak, constant across the substrate and decreasing sharply in intensity within the particle due to absorption and scattering. Spectral factor 4 is a broad feature with an onset around 2 eV and relatively flat intensity across the particle. This factor is observed for all the surveyed structures as well as aggregates and is attributed, due to its spectral shape as well as its spatial distribution, to an interband transition.[60] Spectral factors 5 and 6 are mainly background noise but have been kept to ensure a full analysis.

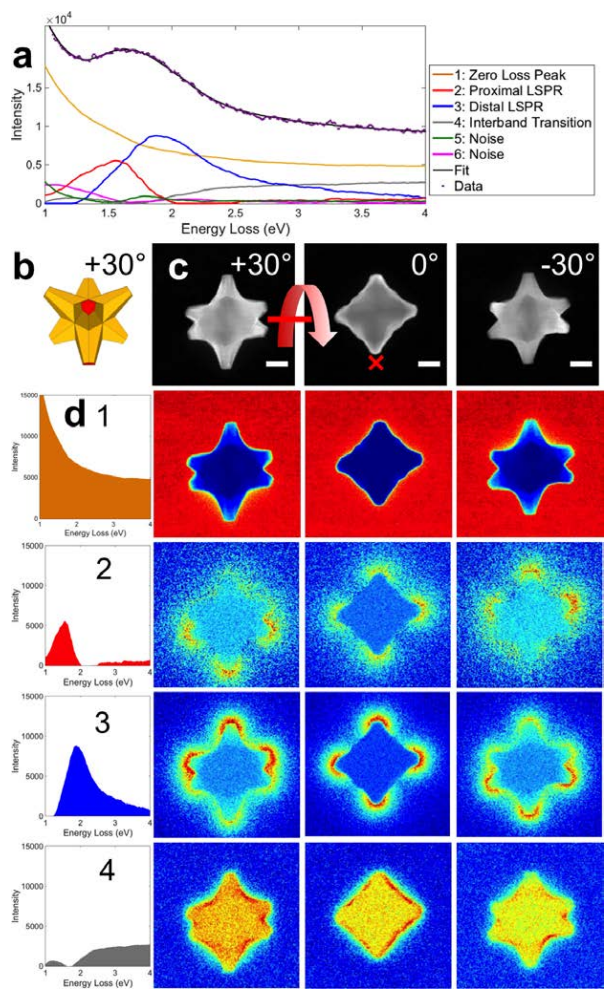

**Figure 3.** Statistical analysis of EELS results on a single Au/Pd octopod. (a) Spectral factors (plasmon modes and other contributions) extracted from NMF and fit of the raw data using the EELS response at the position marked by a "x" in c. (b) Structural model of the 8-branched nanocrystal. (c) Dark field STEM images at different tilts obtained concurrently to the EELS SI. (d) Loadings for the main spectral factors, representing the spatial distribution of (1) the tail of the zero loss peak, (2-3) the low-energy proximal and high-energy distal LSPRs, and (4) interband transitions. Scale bars, 50 nm. The EELS and STEM images have the same scale for each tilt.

While EELS provides a high spatial resolution and ever-improving energy resolution, cathodoluminescence (CL) can provide complementary information that can help better understand complex plasmonic behavior in coupled particle: EELS excites all LSPRs, while CL only detects bright modes. We performed STEM-CL experiments at 200kV with a Gatan Vulcan holder to better understand the dimer coupling, the effect of Pd, and establish the feasibility of such measurements on the Au/Pd system (Figure 4). The panchromatic-CL (all energies acquired) map of a 127 nm octopod clearly shows that Au retains its plasmonic characteristics even in the presence of Pd and that the highest field intensity is located at the Pd-rich tips. The CL spectra obtained at various positions on the nanoparticle peak around 2 eV, as expected for a particle that size; this lowest energy LSPR is a bright dipolar resonance. STEM-CL of an octopod dimer shows the bright bonding LSPR and high field intensity at the tips distant from the gap, this resonance corresponds to the 1.5 eV LSPR in Figure 2d. Moreover, almost no emission is observed from the interparticle gap region, confirming that the LSPR in Figure 2e is indeed a dark antibonding mode and that Pd tips can couple akin their Au counterparts.

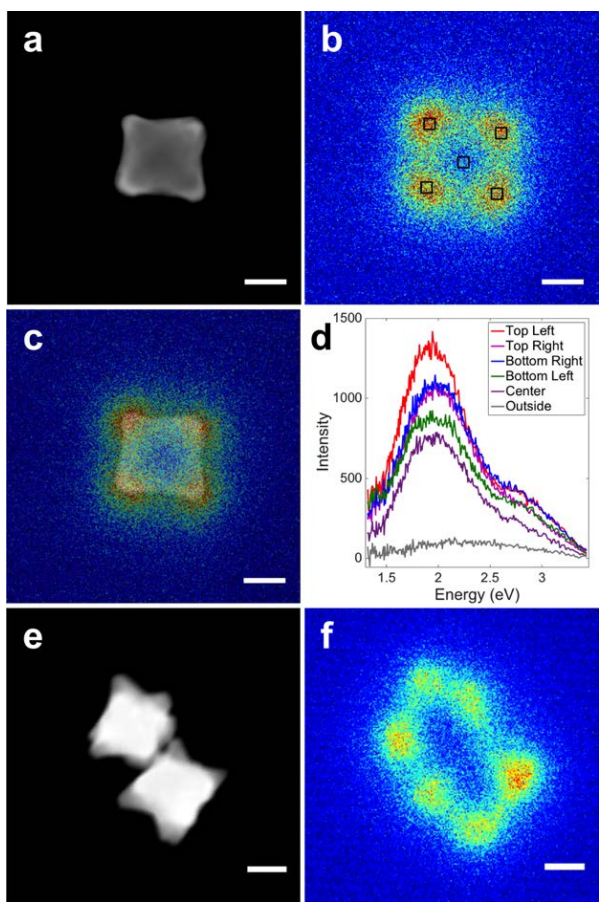

**Figure 4.** STEM-CL spectroscopy of Au/Pd octopods. (a) HAADF-STEM image of a single octopod. (b) Panchromatic-CL image of the octopod in a. (c) Overlay of panchromatic-CL and HAADF-STEM images from a and b. (d) Spectra obtained at the positions marked in b. (e) HAADF-STEM image of an octopod dimer. (f) Panchromatic-CL image of the dimer in f. Scale bars, 50 nm.

Bimetallic nanoparticles with well-defined, controllable geometries are promising multifunctional platforms for various applications including optical sensing and catalysis; amongst such structures are the sharp Au/Pd octopods synthesized *via* a colloidal, seed-mediated co-reduction. In this paper, we have shown that single-crystalline Au/Pd octopods display plasmonic behavior despite the presence of Pd, a metal not suitable on his own for plasmonic applications because of heavy damping. Well-defined, narrow plasmon modes related to those in nanocubes were observed, and their localized near-field intensity was found to sustain strong localized field enhancement at the Pd-rich tips. This observation is extremely encouraging for the field of multifunctional metallic nanoparticles and opens the door for further studies of bimetallic nanoparticles, where the plasmonic core can provide not only sensing, but well-defined local field enhancements co-localized with catalytically active materials.

ACKNOWLEDGMENT

The authors acknowledge financial support from the European Union under the Seventh Framework Programme under a contract for an Integrated Infrastructure Initiative (Reference 312483 - ESTEEM2) and from the ERC grant 3DIMAGE (291522). E. R. acknowledges support from the Royal Society (Newton International Fellowship) and Trinity Hall Cambridge. S. M. C. acknowledges support from a Gates Fellowship. S. E. S. and C. J. D. acknowledge support from NSF CHE 1306853.

ASSOCIATED CONTENT

**Supporting Information**. Nanoparticle synthesis details, tomography snapshots and tilt series, additional diffraction and high resolution images, additional EDS, EELS, and CL data and experimental details. This material is available free of charge via the Internet at http://pubs.acs.org.

AUTHOR INFORMATION

**Corresponding Author**

*Emilieringe@rice.edu

# Supplementary Information

# Near-Field Plasmonic Behavior of Au/Pd Nanocrystals with Pd-Rich Tips

*Emilie Ringe,[1*] Christopher J. DeSantis,[2] Sean M. Collins,[3] Martial Duchamp,[4] Rafal E. Dunin-Borkowski,[4] Sara E. Skrabalak,[2] Paul A. Midgley[3]*

1. Department of Materials Science and NanoEngineering, Rice University, 6100 Main St., Houston TX 77005, USA

2. Department of Chemistry, Indiana University, 800 E. Kirkwood Ave., Bloomington, IN 47405, USA

3. Department of Materials Science and Metallurgy, University of Cambridge, 27 Charles Babbage Road, Cambridge CB3 0FS, UK

4. Ernst Ruska-Centre for Microscopy and Spectroscopy with Electrons (ER-C) and Peter Grünberg Institut 5 (PGI-5), Forschungszentrum Jülich GmbH, D-52425 Jülich, Germany

## NANOPARTICLE SYNTHESIS

Au/Pd octopods were synthesized according to a previously reported method.[1] Au nanoparticles cores were obtained from a seed-mediated reduction of $HAuCl_4$ by L-ascorbic acid in the presence of cetyltrimethylammonium bromide (CTAB). Small cores were then coated with a Au/Pd alloy through co-reduction of $H_2PdCl_4$ and $HAuCl_4$ by L-ascorbic acid at room temperature. Figure S1 shows a representative sample with an average tip-to-tip (face diagonal) distance of 118 (±8) nm and tip thickness of 21 (±3) nm. Most of the particles in the reaction mixture were octopods, however some larger star-shaped particles derived from right bipyramids were present; these structures are rare and synthetic advances have reduced their occurrence.[2,3] Particles were drop cast from solution on $Si_3N_4$ membrane windows (EELS, EDS, HR-TEM, diffraction, Tomography; TEMWindows.com), or carbon-coated grids (CL, Tomography; Pacific Grid Tech).

SHAPE AND SIZE CHARACTERIZATION

Tomographic reconstructions (Figure S1) were obtained from HAADF-STEM tilt series acquired (Figure S2) on a FEI Tecnai FEG operated at 200 kV using a simultaneous iterative reconstruction technique (SIRT) in the software Inspect 3D (FEI Company). The voxel projection visualization of reconstructed volume was obtained in Avizo Fire (Visualization Science Group), without any cropping of the area surrounding the particle.

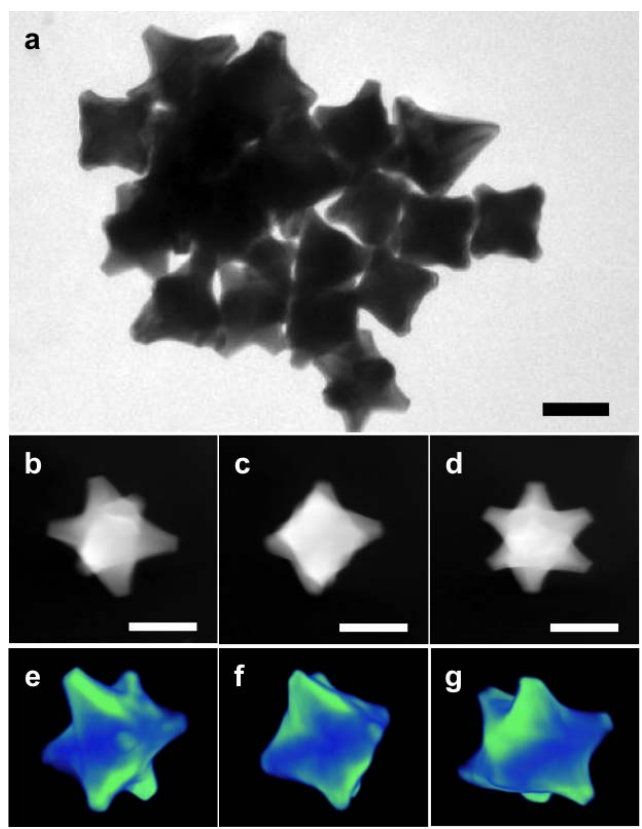

**Figure S1.** Structure of Au/Pd octopods. (a) Bright field TEM. (b-d) HAADF-STEM images of an octopod at -50°, 0°, and +50°. (e-g) Snapshots of the 3-dimensional reconstruction obtained from electron tomography of the particle shown in b-d. Scale bars, 100 nm.

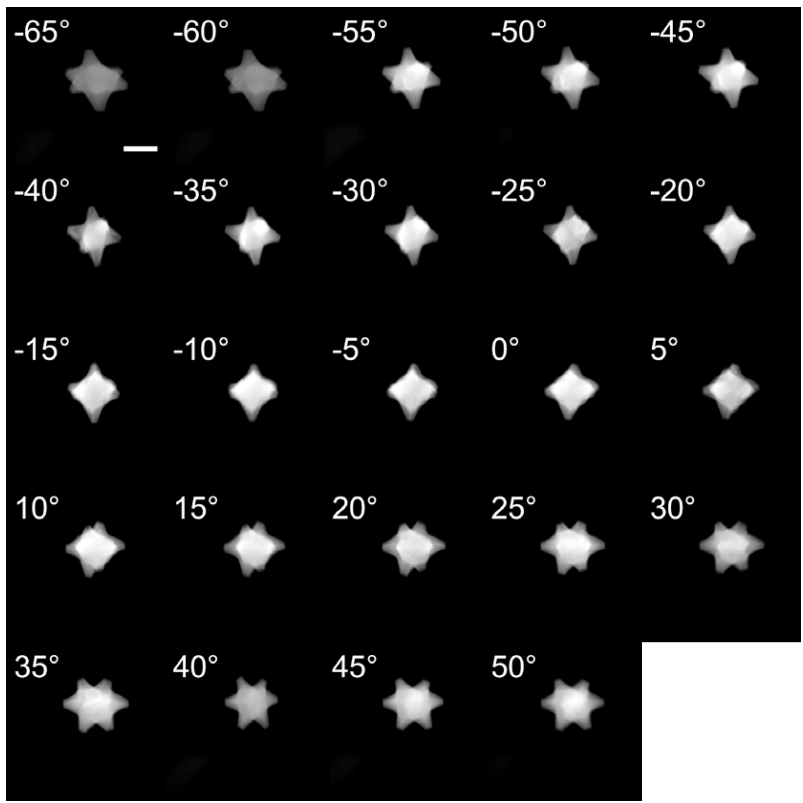

**Figure S2.** HAADF-STEM tilt series for the octopod shown in Figure S1. Scale bar, 100nm; the scale is the same for all the pictures.

CRYSTALLOGRAPHY AND DIFFRACTION

Electron diffraction and high resolution imaging (Figures 1, S3c-d, S5a-c) were performed in a probe-corrected JEOL ARM CFEG operated at 200 kV. Diffraction patterns taken with the beam perpendicular to one of the underlying cube faces all show the prominent 4-fold symmetry attributable to a FCC <100> orientation. Diffraction mapping shows that every diffraction pattern has the same <100> symmetry and orientation, confirming the single crystalline nature of the octopod, as was previously observed for FIB-cut nanocrystals.[4]

The lattice spacing difference between Au and Pd is of the order of the measurement error, such that using it for compositional analysis is difficult. Every lattice spacings measured, either via FFT, atomic column measurements, or diffraction patterns, fell between the value for Au and that for Pd.

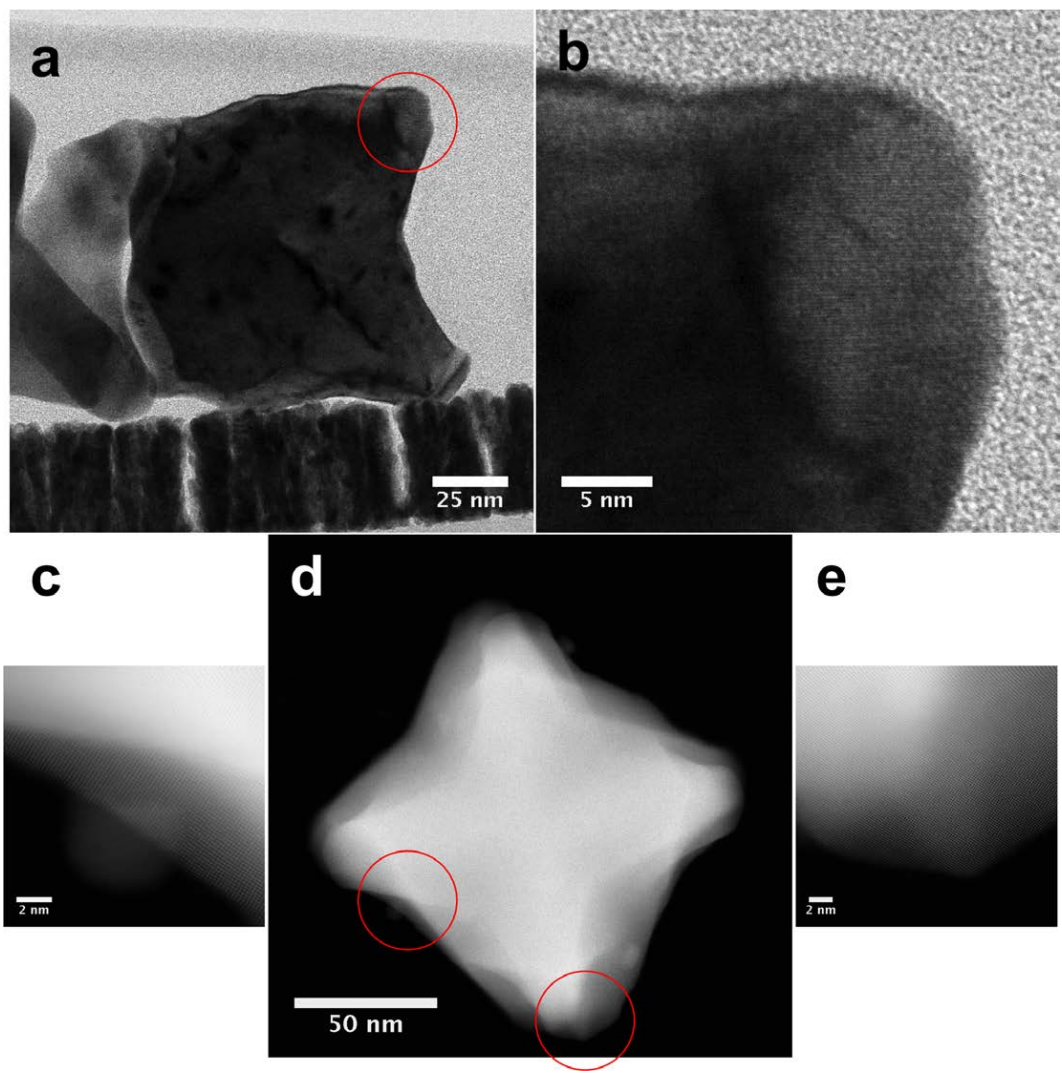

**Figure S3.** Additional atomic resolution images of octopods highlighting the absence of defects and the single crystalline nature of the branches. (a) Bright field STEM image of an octopod cut in half by an FIB, adapted with permission from ref.[4] (b) Higher magnification of the region shown in a, the horizontal lattice fringes correspond to {200} planes with a distance of ~0.2 nm. (d) HAADF-STEM images of an octopod. (c, e) Higher magnification of the regions marked in d.

ENERGY DISPERSIVE X-RAY SPECTROSCOPY (EDS) ANALYSIS

All EDS spectra except those of Figure S6 were acquired in a FEI Titan Chemistem operated at 200 kV, using a Bruker Super-X quad EDS detector. The EDS analysis (except Figure S6) was performed in HYPERSPY. The relative X-ray intensity was obtained by first subtracting the background, then integrating the peak intensity over an appropriate energy range, and finally dividing the integrated peak intensity by the total X-ray count. This process is necessary to separate the effects of thickness and composition, as shown in Figure S4, where the non-normalized intensity profiles (Figure S4b) are dominated by thickness variation, where the two broad peaks (~100 and ~180 nm) correspond to the position of two branches, one above and one below the particle core. The normalized linescans (Figures 1j and S4c) make it clear that Pd is present in greater relative amount at the tips, and that the stoichiometry within the particle is relatively constant.

To produce the images presented in Figures 1 and S5, areas outside the particle were set to zero relative intensity by applying an intensity threshold for the X-ray peak plotted; this was necessary in order to avoid large fluctuations due to the division of two small noise-dominated numbers. Figure S6 shows the net integrated signal of the Au $M_\alpha$ and $L_\alpha$ and Pd $L_\alpha$ and $L_\beta$ data acquired on a JEOL ARM Cold FEG operated at 200 kV, as an additional and higher resolution confirmation of the presence of Pd at the sides and tips of the particle. The octopod analyzed in Figure S6 is the same as in the bottom of Figure S3.

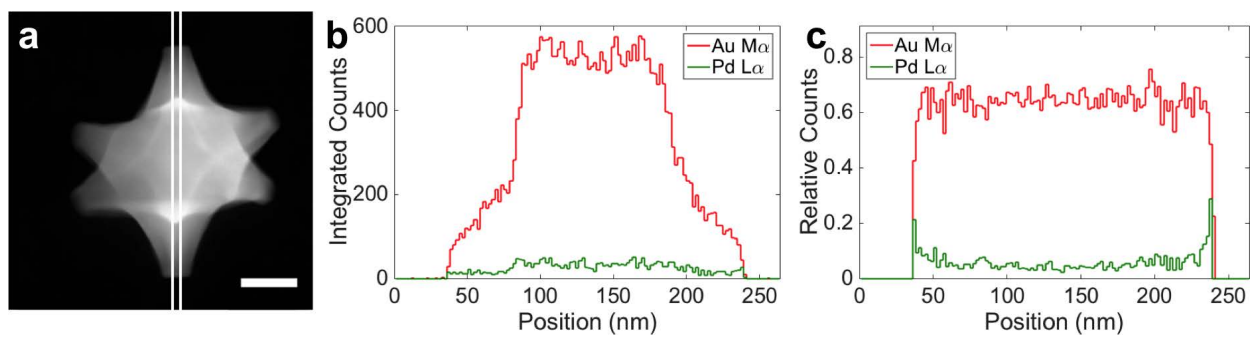

**Figure S4.** Additional EDS linescan data for a single octopod tilted -30°. (a) HAADF-STEM image. (b) Background subtracted, integrated counts for Au $M_\alpha$ and Pd $L_\alpha$ lines. (c) Background subtracted, integrated and normalized counts for Au $M_\alpha$ and Pd $L_\alpha$ lines. Scale bars, 50 nm.

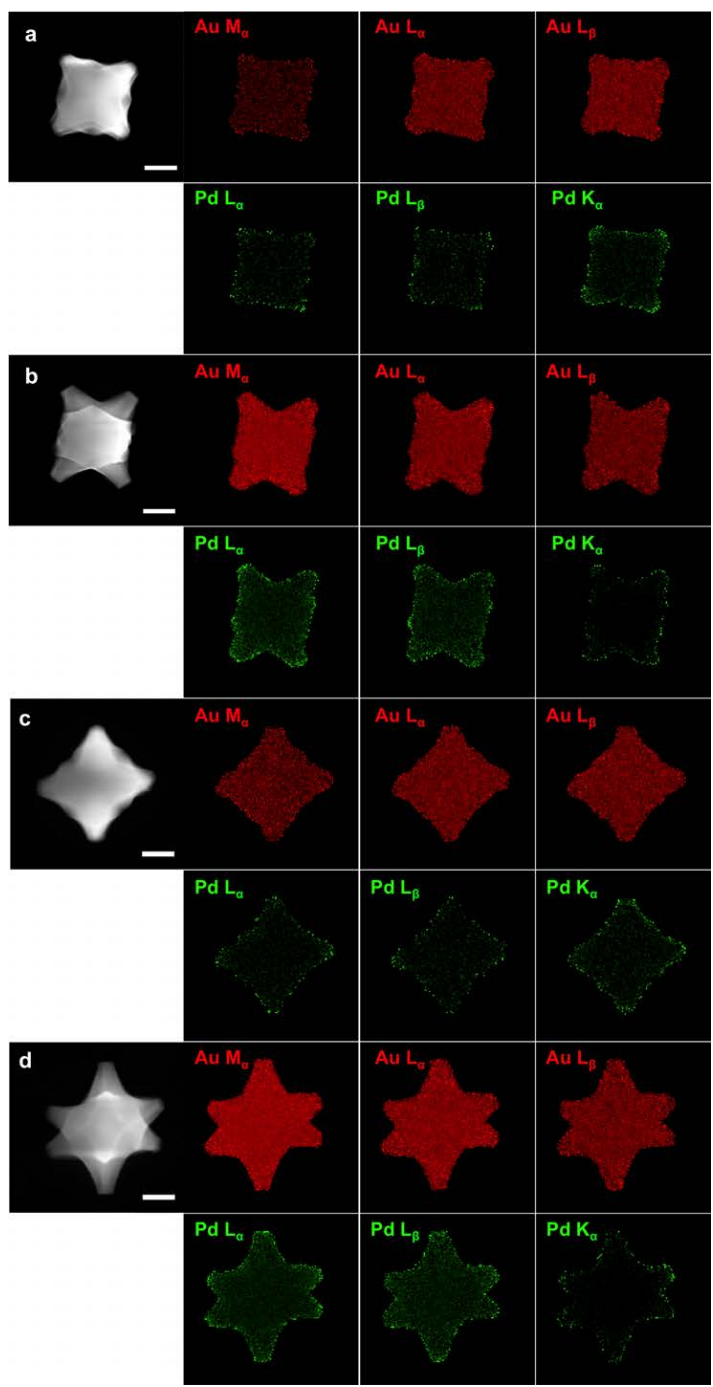

**Figure S5.** HAADF-STEM and EDS maps of two octopods. (a-b) Octopod in Figure S8, tilted 0° and +30°. (c-d) Octopod in Figures 2a, 3, S6-S7, tilted 0° and -30°. Data processing as explained in the text. Scale bars, 50 nm. The EDS and STEM images have the same scale.

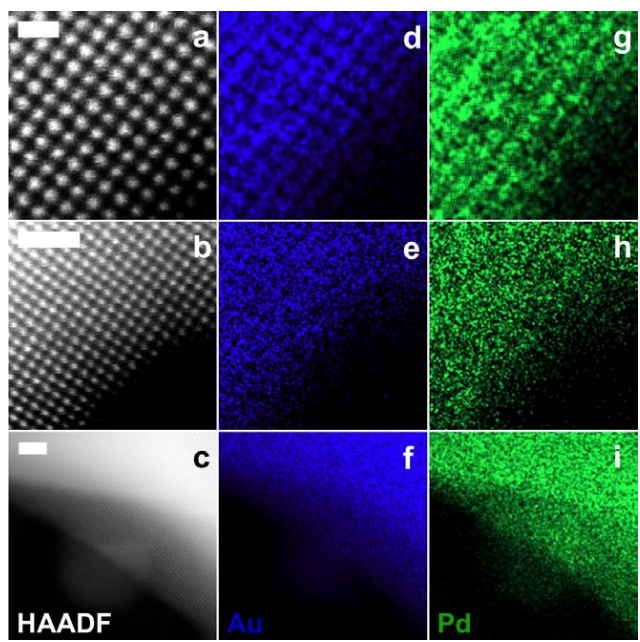

**Figure S6.** EDS analysis of Au/Pd octopods performed on a JEOL ARM. (a-c) HAADF-STEM images. (d-e) Au intensity from the $M_\alpha$ and $L_\alpha$ line. (g-h) Pd intensity from the $L_\alpha$ and $L_\beta$ lines. Scale bars, 0.5 nm for a, 1 nm for b, 2 nm for c. The EDS and STEM images have the same scale.

## PLASMON MAPPING WITH EELS

STEM-EELS (Figures 2a-3, S6-8) was performed on a probe corrected, monochromated FEI Titan Themis operated at 200 kV, equipped with an X-FEG electron gun and a Wien filter monochromator, or (Figure 2d-e) a JEOL ARM Cold FEG operated at 200 kV. EELS spectra were acquired with a Gatan GIF Quantum ERS energy-loss spectrometer (Figures 2a-3, S6-8) or a Gatan Enfinium ER energy-loss spectrometer (Figure 2d-e).

Spectrum images (SI) measuring 164X164 pixels we acquired at +30°, 0°, and -30° for the particle presented in Figures 2a-3 and S6-S7. A SI measuring 128X207 pixels was acquired at 0° for the dimer in Figure 2d-e.

Simulated energy-filtered TEM (EFTEM) images were obtained by integrating the STEM-EELS intensity over a 0.2 eV window centered at 1.5 and 2.3 eV in Figure 2 and 1.65 and 2.5 eV in Figure S6. The EFTEM energy values were selected by determining the apparent peak position in the EEL spectra. Simulated EFTEM images at high energy (>5 eV) closely resemble the spatial distribution of the interband transitions, however it is difficult to assess the interband nature as the spectral shape is not provided from simulated EFTEM (it is from NMF).

Note that the "As-aquired" SI and HAADF-STEM data at 0° in S8 is rotated 45° with respect to the tilted SI (+30° and -30° tilt). The SI and HAADF (164X166 pixels) were collected at 0° prior to a 45° rotation of the sample. For ease of readability, the HAADF-STEM and loading maps in the third column of Figure S8 were reproduced from the second column with the appropriate 45° image rotation and crop to match the orientation of the +30° (142X142 pixels) and -30° (164X164 pixels) SI subsequently acquired.

The multidimensional data arrays (164X164X2048X3 — 164X164 region of interest, 2048 energy channels, 3 tilts for the particle in Figures 2a-3, S7 and 142X142X2048X3 for the particle in Figure S8) were analyzed using blind source separation of decomposed modes from non-negative matrix factorization performed in HYPERSPY.[5,6] This approach decomposes the intrinsically redundant information of a spectrum image (SI) into a number of spectral components (spectral factors) that are multiplied by different coefficients (loadings) at each pixel to best fit the SI. The factors were not assigned fixed peak shapes. The spectra analyzed were cropped from 0.3 to 5 eV and all three tilts for a given particles were processed simultaneously.

The two distinct particles in Figures S7 and S8 were present on the same $Si_3N_4$ grid and investigated on a monochromated FEI Titan Themis in the course of a single experiment, making the tilting behavior (axis orientation and tilt direction) directly comparable.

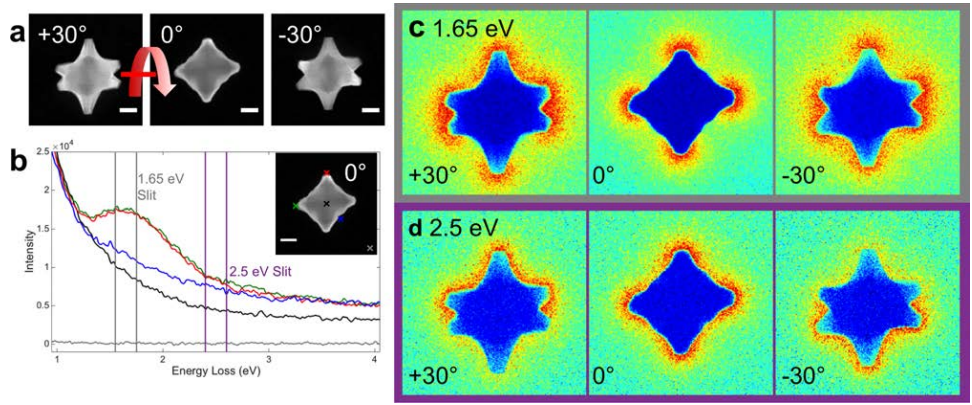

**Figure S6.** Simulated EFTEM images. (a) HAADF-STEM images at various tilts, the horizontal red line indicates the tilt axis. (b) Raw data integrated over 11X11 pixels at the locations indicated in the inset, as in Figure 4. (c-e) Simulated EFTEM images obtained from integrating the intensity using a virtual 0.2 eV slit centered at the apparent peak position, 1.65 eV, as well as the tail of the plasmon, 2.5 eV. The 2.5 eV data is consistent with the conclusion that the plasmon is hybridized with the substrate in a proximal and distal configuration. Scale bar, 50 nm. The EELS and STEM images have the same scale.

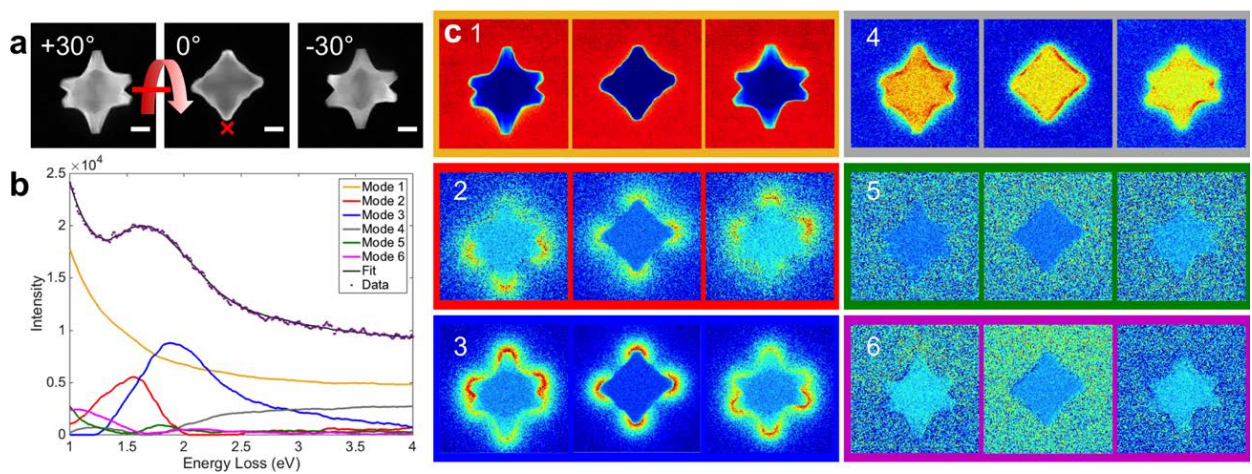

**Figure S7.** Full statistical analysis of EELS results on a single Au/Pd octopod, the same particle shown in Figure 3. (a) Dark field STEM images obtained concurrently to the EELS SI. (b) Spectral factors extracted from NMF as well as the fit of the raw data using the EELS response at the position marked in a. (c) Loadings for each of the spectral factors (1-6), representing the intensity of each factor in b at every pixel of the image. Spectral factor 1 is the tail of the ZLP, 2 is the low-energy proximal (into the substrate) LSPR, 3 is the high-energy distal (away from the substrate) LSPR, 4 is due to interband transitions, 5 and 6 are noise included to insure a full and complete analysis. Scale bars, 50 nm. The EELS and STEM images have the same scale for each tilt.

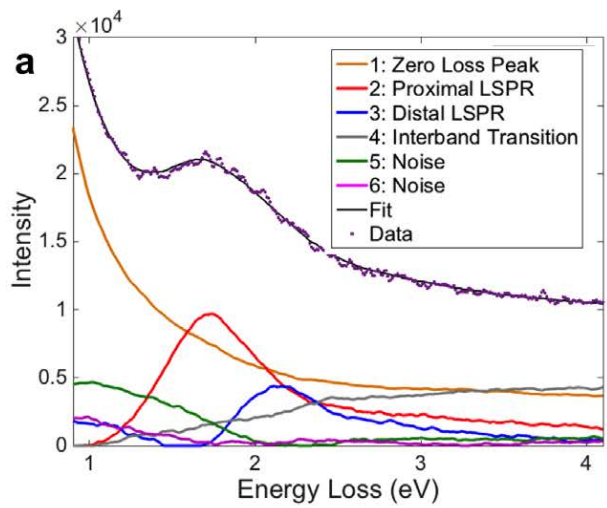
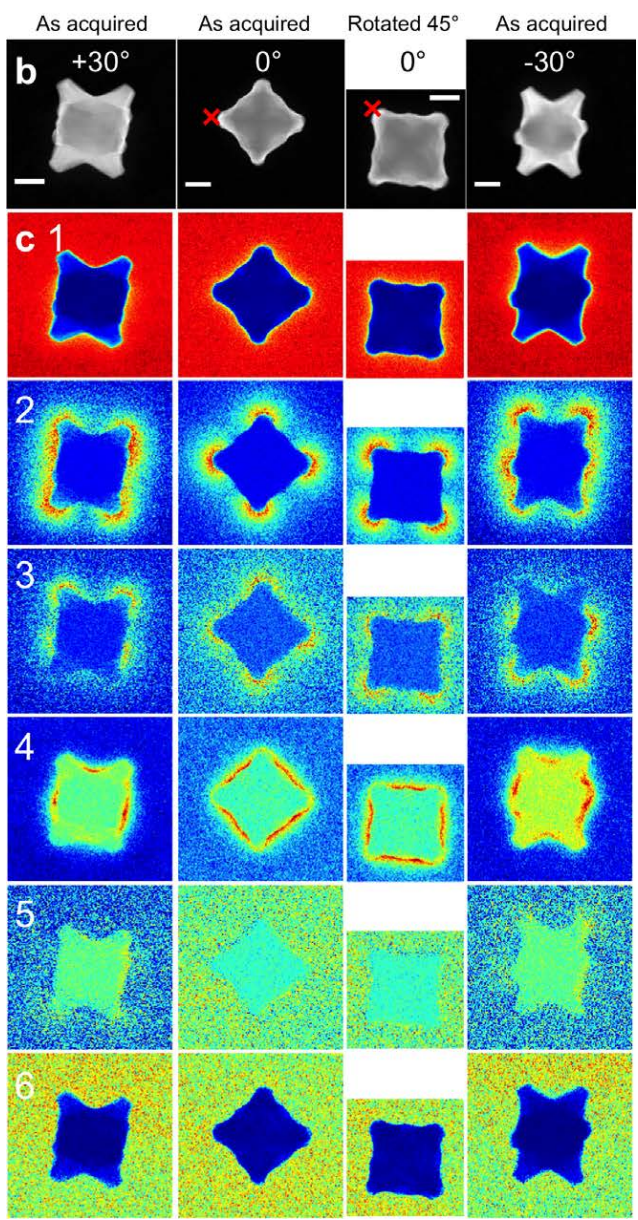

**Figure S8.** Full results of NMF for an additional single octopod rotated approximately 45° from the octopod in Figures 2a-3 and S6-7. (a) Spectral factors extracted from NMF as well as the fit of the raw data using the EELS response at the position marked in b. (b) Dark field STEM images obtained concurrently to the EELS SI. (c) Loadings for each of the spectral factors (1-6), representing the intensity of each factor in b at every pixel of the image. Spectral factor 1 is the tail of the ZLP, 2 is the low-energy proximal (into the substrate) LSPR, 3 is the high-energy distal (away from the substrate) LSPR, 4 is due to interband transitions, 5 and 6 are noise and remaining ZLP contributions included to insure a full and complete analysis. The field intensity distribution orientation for the LSPR modes matches well that of Figure 3. The first, second, and fourth columns in both b and c are "as acquired", while the third column shows data rotated 45° to match the orientation of the first and fourth columns. Scale bars, 50 nm; the EELS and STEM images have the same scale.

CATHODOLUMINESCENCE

Cathodoluminescence measurements reported in Figures 4 and S9 were performed using a Vulcan holder (GATAN, Inc) inserted into a JEOL 2100F STEM operated at 200 kV.

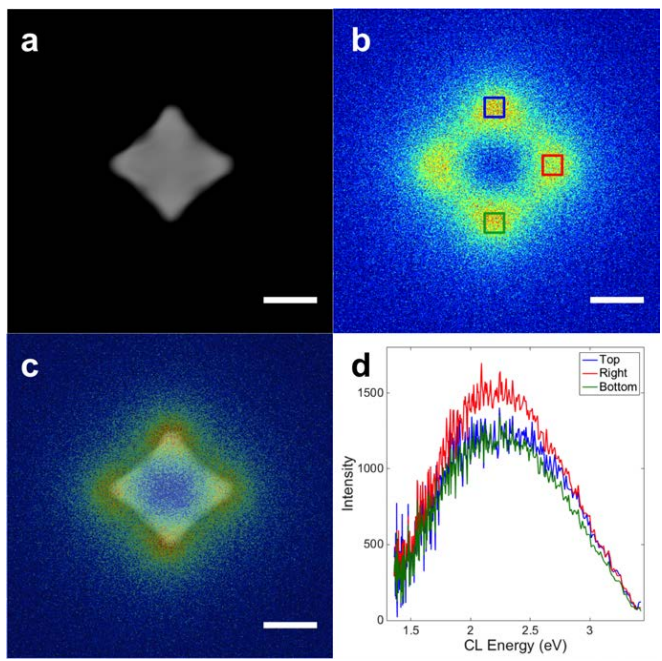

**Figure S9.** Additional STEM-CL data from of a single Au/Pd octopod measuring 100 nm tip to tip. (a) HAADF-STEM image, (b) Panchromatic-CL image. (c) Overlay of panchromatic-CL and HAADF-STEM images. (d) Spectra obtained at the positions marked in b. Scale bars, 50 nm.